\documentclass[12pt]{article}
\newcommand\unit[1]{{\;\rm #1}}
\def\blambda{\lambda\!\!\!\raise.55ex\hbox{$\,${-}}\,}

\newcommand\pd[2]{\frac{\partial#1}{\partial#2}}
\begin{document}
\thispagestyle{empty}
\vspace*{2cm}
\begin{center}
{\huge  Beyond the Fourier equation:\\
Quantum hyperbolic heat transport}\\
\vspace{24pt}
Miroslaw Kozlowski$^*$\\
Institute of Experimental Physics\\
Warsaw University\\
Ho\.za 69, 00-681 Warsaw, Poland\\
 e-mail: mirkoz@fuw.edu.pl

\bigskip

Janina Marciak-Kozlowska\\
Institute of Electron Technology\\
Al.~Lotnik\'ow~32/46, 02--668~Warsaw, Poland
\end{center}

\bigskip\bigskip

\begin{abstract}
In this paper the quantum limit of heat transport induced by ultrashort laser
pulses is discussed. The new quantum heat transport equation is derived. The relaxation
time $\tau=\hbar/mv_h^2$ ($v_h$ = thermal wave velocity) and diffusion
coefficient $D^e=\hbar/m$ are calculated.
\end{abstract}

\bigskip


\bigskip
\bigskip

\hbox to 5cm{\hsize=5cm\vbox{\ \hrule}}\par
\noindent{\llap{$^{\rm *}$~}Author to whom correspondence should be addressed.}

\newpage
\section{Introduction}

The correlated random walk, whose diffusive analog is described by hyperbolic
diffusion equation, is possibly the simplest mathematical model allowing one to
incorporate a form of momentum in addition to random or diffusive motion. The
correlated random walk differs from the ordinary random walk in that the
probabilistic element used at each step is the probability of continuing
to move in
a given direction rather than the probability of moving in a given direction
independent of the direction of the immediately preceding step. Thus the
process remains a Markov process.

The solution to a variety of forms of the hyperbolic diffusion
equation has been given by Goldstein~\cite{1}.

There is considerable literature~\cite{2, 3, 4} on physical
processes in the field of thermophysics leading to a mathematical
formulation in terms of a hyperbolic heat conduction equation. The
physics behind such application implies that the signal
propagation speed is finite rather than infinite as in the case of
ordinary diffusion.

Until recently the thermal phenomena studied with the help of
hyperbolic heat transfer equation do not consider the quantum
aspects of heat transfer~\cite{2, 3, 4}. All the results are
obtained in classical approximations.

In this paper the quantum limit of heat transfer will be
considered and quantum heat transport equation will be developed.
The study was motivated by the advent of the ultra~--~short
duration laser pulses, particularly those in attosecond
($10^{-18}$)~\cite{5} and zeptosecond~\cite{6} domains.
Considering that time in which electron orbits the atomic nucleus
is of the order of femtoseconds~---~with attosecond pulses it is
possible penetrate the quantum structure of the path of electrons.

\section{Overview of research}
In present day materials sciences laser technology is a fundamental tool in
such processes as surface deformation, materials hardening, welding, etc.
Recently it has been shown possible to study accompanying phenomena on an
atomic level. It is well recognized that on deposition of the laser pulse,
electromagnetic energy is absorbed by a~photon~--~electron interaction, the
magnitude of which is described by the fine structure constant $\alpha=1/137$.
The response of material to the thermal perturbation is governed by the
relaxation time $\tau$. These two parameters fully describe the dynamics of
energy transfer of the ultra~--~short (femtosecond and shorter) laser pulses on
a nanometer scale. Conditions in the processes induced by ultra~--~short laser
pulses are in stark contrast with those of longer laser pulse duration, the
distinction being that at longer pulse duration ($\sim$ 1~ps) excited particles
and their surroundings have had sufficient time to approach thermal equilibrium.
On the other hand at temporal resolution $\sim$ 1~fs it is possible to resolve the
dynamics of nonstationary transport phenomena.

For the transport phenomena of the nonrelativistic electrons there is a
stochastic process that is ultimately connected with its propagation, namely
the Brownian motion. As is well known Nelson~\cite{7}
succeed in deriving the Schr\"{o}dinger equation from the assumption that
quantum particles follow continous trajectories in a chaotic background. The
derivation of the usual Schr\"{o}dinger equation follows only if the diffusion
coefficient {\it D} associated with quantum Brownian motion takes the value
$D=\hbar/{2m}$ as assumed by Nelson.

In this paper the diffusion process of the quantum particles in context of the
thermal energy will be studied. In the paper~\cite{8}
 the heat transport
in a thin metal film (Au) was investigated with the help of the
hyperbolic heat conduction equation (HHC). It was shown that when
the memory of the hot electron gas is taken into account then the
HHC is the dominant equation for heat transfer. The hyperbolic
heat conduction equation for heat transfer in an electron gas can
be written in the form~\cite{8}:
\begin{equation}
\frac{1}{c^2}\pd{^2T}{t^2}+\frac1{c^2\tau}\pd{T}{t}=\frac{\alpha^2}3\nabla^2T.\label{eq0}
\end{equation}
In equation (\ref{eq0}) $c$ is the light velocity in vacuum,
$\alpha=e^2\hbar^{-1}c^{-1}=1/137$ is the fine structure constant and $\tau$ is
the relaxation time.

In the present paper the quantum limit of the equation~(\ref{eq0}) will be
obtained. When a~high energy laser pulse hits the thin metal film it is possible
that very short relaxation time are obtained. Since the relaxation time
$\tau$ strongly depends on the temperature the relaxation times are of the
order of 1 fs for $T>10^3$ K. In this case the mean free path of the hot
electron is the same order as the thermal de Broglie wavelength. The
temperature region for which mean free path $\lambda$ is of the order of
$\blambda_B$ (de Broglie wavelength) we will define as the quantum limit of the
heat transport and the master equation~---~the quantum heat transport equation
(QHT).

\section{Quantum limit of the heat transport}
The classical notion of the field was born from a description of
material media: the propagation of certain perturbations in such a
medium led to the motion of a wave. A~field, or a wave, gave
initially, therefore, a picture of the collective motions of the
medium and their propagation. For example the fluctuations of the
electronic density in a solid state plasma are described by
``plasma'' waves. The classical formalism of the theory of fields
is the same whether it is applied to ``true'' fields, which are
thought of as existing in themselves, even in the absence of any
material substratum, such as the electromagnetic field, or to a
``phenomenological field'' produced by the collective motions of a
propagating medium. The concept of a frequency, for example, plays
the same fundamental role in the analysis of these fields. The
quantum synthesis of the concepts of frequency and energy is thus
universal, i. e. they do not depend upon the specific nature of
the phenomenon being considered and characterize the quantum
theory in all its generality. The Planck~--~Einstein relation
$$
E=\hbar\omega
$$
holds equally well, therefore, in quantum description of
``phenomenological'' fields. In other words, one must have a
discretization of the energy into packets or quanta,
$E=\hbar\omega$, for any waves: acoustic or thermal, with a
vibration frequency $\omega$, exactly as for an electromagnetic
wave.

In~\cite{9} we developed the new hyperbolic heat transport
equation which generalizes the Fourier heat transport equation for
the rapid thermal processes. The hyperbolic heat conduction
equation for the fermionic system can be written in the form:
\begin{equation}
\frac1{\left(\frac13v_F^2\right)}\;\pd{^2T}{t^2}+
\frac1{\tau\left(\frac13v_F^2\right)}\;\pd
Tt=\nabla^2T,\label{eq1}
\end{equation}
where $T$ --- denotes the temperature, $\tau$ --- the relaxation time for the
thermal disturbance of the fermionic system and~$v_F$ is the Fermi velocity.

In the subsequent we develop the new formulation of the HHC
considering the details of the fermionic system.

To start with we recapitulate the well known results from the simplest
fermionic model of the matter: the Fermi gas of electrons in metals.

For the electron gas in metals the Fermi energy has the form~\cite{10}:
\begin{equation}
E_F^e=(3\pi^2)^{2/3}\,\frac{n^{2/3}\hbar^2}{2m_e},\label{eq2}
\end{equation}
where $n$ --- density and $m_e$ --- electron mass. Considering that
\begin{equation}
n^{-1/3}\sim a_B\sim\frac{\hbar^2}{m_ee^2}\label{eq3}
\end{equation}
and $a_B={}$Bohr radius, one obtains
\begin{equation}
E_F^e\sim\frac{n^{2/3}\hbar^2}{m_e}\sim\frac{\hbar^2}{ma^2}\sim\alpha^2m_ec^2,
\label{eq4}
\end{equation}
where $c={}$light velocity and $\alpha=1/137$ is the fine
structure constant. For the Fermi momentum $p_F$, we have
\begin{equation}
p_F^e\sim\frac{\hbar}{a_B}\sim\alpha m_ec\label{eq5}
\end{equation}
and for Fermi velocity~$v_F$,
\begin{equation}
v_F^e\sim\frac{p_F}{m_e}\sim\alpha c.\label{eq6}
\end{equation}
Considering formula~(\ref{eq6}), equation~(\ref{eq1}) can be
written as
\begin{equation}
\frac1{c^2}\;\pd{^2T}{t^2}+\frac1{c^2\tau}\;\pd
Tt=\frac{\alpha^2}3\,\nabla^2T.\label{eq7}
\end{equation}
As it is seen from~(\ref{eq7}) the~HHC equation is the relativistic equation as
it takes into account the finite velocity of light. In order to derive the
Fourier law from equation~(\ref{eq7}) we are forced to break the special theory
of relativity and put in equation~(\ref{eq7}) $c\to\infty$, $\tau\to0$. In
addition it was demonstrated from HHC in a~natural way, that in electron gas
the heat propagation velocity $v_h\sim v_F$ in the accordance with the results
of the laser pump probe experiments.

In the following the procedure for the discretization of temperature~$T(\vec
r,t)$ in hot fermion gas will be developed. First of all we introduce the
reduced de~Broglie wavelength
\begin{equation}
\blambda_B^e=\frac\hbar{m_ev_h^e}\qquad v_h^e=\frac1{\sqrt3}\,\alpha c\label{eq15}
\end{equation}
and mean free path $\lambda^e$,
\begin{equation}
\lambda^e=v_h^e\tau^e.\label{eq16}
\end{equation}
Considering formulae~(\ref{eq15}), (\ref{eq16}) we obtain HHC for
electron
\begin{equation}
\frac{\blambda_B^e}{v_h^e}\;\pd{^2T^e}{t^2}+\frac{\blambda_B^e}{\lambda^e}\;
\pd{T^e}{t}=\frac\hbar{m_e}\,\nabla^2T^e.\label{eq17}
\end{equation}
Equation~(\ref{eq17}) is the hyperbolic partial differential
equation which is the master equation for heat propagation in Fermi electron
 gas. In the following we will study the quantum limit of heat
transport in the fermionic systems. We define the quantum heat transport limit
as follows:
\begin{equation}
\lambda^e=\blambda_B^e.\label{eq19}
\end{equation}
In that case equation~(\ref{eq17}) has the form:
\begin{equation}
\tau^e\,\pd{^2T^e}{t^2}+\pd{T^e}t=\frac\hbar{m_e}\,\nabla^2T^e,\label{eq20}
\end{equation}
where
\begin{equation}
\tau^e=\frac\hbar{m_e(v_h^e)^2}.\label{eq22}
\end{equation}
Equation~(\ref{eq20}) defines the master equation for quantum heat
transport~(QHT). Having the relaxation time~$\tau^e$ one can
define the ``pulsations''~$\omega_h^e$:
\begin{equation}
\omega_h^e=(\tau^e)^{-1}\label{eq23}
\end{equation}
or
\[\omega_h^e=\frac{m_e(v_h^e)^2}\hbar\],
i.e.
\begin{equation}
\omega_h^e\hbar=m_e(v_h^e)^2=\frac{m_e\alpha^2}3\,c^2.\label{eq24}
\end{equation}
Formula~(\ref{eq24}) defines the Planck-Einstein relation for heat
quanta~$E_h^e$
\begin{equation}
E_h^e=\omega_h^e\hbar=m_e(v_h^e)^2.\label{eq25}
\end{equation}
The heat quantum with energy~$E_h=\hbar\omega$ can be named as the
{\it heaton\/} in complete analogy to the {\it phonon, magnon,
roton\/} and etc. For $\tau^e \to 0$ equations~(\ref{eq20}),
(\ref{eq24}) are the quantum Fourier transport equations (QFT)
with quantum diffusion coefficients $D^e$:
\begin{equation}
\pd{T^e}t=D^e\nabla^2T^e,\qquad D^e=\frac\hbar{m_e}.\label{eq26}
\end{equation}
For finite $\tau^e$ for $\Delta t<\tau^e$ equation~(\ref{eq20})
can be written as follows:
\begin{equation}
\frac1{(v_h^e)^2}\;\pd{^2T^e}{t^2}=\nabla^2T^e.\label{eq28}
\end{equation}
Equation~(\ref{eq28}) is the wave equation for quantum heat
transport~(QHT).

It is interesting to observe that equation~(\ref{eq28}) is the
wave equation which describes the propagation of the de Broglie
thermal wave. For the wave length $\blambda$ of the wave which is
the solution of equation~(\ref{eq28}) one obtains:
\begin{equation}
\blambda=\frac{v_h}{\omega}=\frac{v_h\hbar}{\omega\hbar}=\frac{\hbar}{p_e},\qquad
p_e=m_ev_h,\label{eq29}
\end{equation}
i.e. $\blambda$ is equal the reduced de Broglie wave length~(\ref{eq15}).

On the other hand the quantum Fourier equation~(\ref{eq26}) resembles the free
Schr\"{o}dinger equation. The replacement $t\rightarrow it/2$ turns the quantum
diffusion equation into the Schr\"{o}dinger equation. Both are parabolic and
require the same boundary and initial conditions in order to be ``well posed''.

The {\it heaton\/} energies for electron  gas can be
calculated from formula~(\ref{eq25}). For electron gas we
obtain from formulae~(\ref{eq15}), (\ref{eq25})
for~$m_e=0.51\unit{MeV/{\mit
c}^2}$, $v_h=(1/\sqrt{3})\alpha c$
\begin{equation}
E_h^e=9\unit{eV},\label{eq30}
\end{equation}
which is of the order of the Rydberg energy. The numerical values of the
relaxation time $\tau$ and the quantum diffusion coefficient $D$ can be
calculated from formulae~(\ref{eq22}) and~(\ref{eq26}).
$$
\tau=10^{-17}\unit{s}, \qquad D=1.2\,10^{-4}\unit{m^2s^{-1}}.
$$
Concluding, one can say that for the temporal resolution $\Delta
t$ of the order or shorter of the relaxation time $\tau \sim
10^{-17}\unit{s}$ the heat transport phenomena are adequately
described by the quantum heat transport equation (QHT),
formula~(\ref{eq20}). It seems that the advent of lasers with
attosecond laser pulses open quite new possibility for the study
theses discrete thermal phenomena~\cite{10}.


\newpage
\begin{thebibliography}{99}
\bibitem{1}S.~Goldstein, On Diffusion by Discontinuous Movements and on
 the Telegraph Equation,  \textit{Quart. Journ. Mech. Applied Math.}, \textbf{vol.~IV}, 1951, pp.
129--156.
\bibitem{2}D.~D.~Joseph, L.~Preziosi, Heat Waves, \textit{Rev. Mod. Phys.},
\textbf{vol.~6}, 1989, pp.41--73.
\bibitem{3}D.~Jou et al., \textit{Extended Irreversible
Thermodynamics}, Springer, Berlin, 1996.
\bibitem{4}I.~M\"{u}ller,  T.~Ruggeri, \textit{Extended  Thermodynamics},
Springer, Berlin, 1993.
\bibitem{5}M.~Ivanow et al., Routes to Control of Intense  Field Atomic
Polarizability, \textit{Phys. Rev. Lett.}, \textbf{vol.~74}, 1995,
pp.2933--2936.
\bibitem{6}A.~E.~Kaplan, P.~L.~Shkolnikov, \textit{Phys. Rev.
Lett.}, \textbf{88}, 2002, p.~074801-1.
\bibitem{7}E.~Nelson, Derivation of the Schr\"odinger Equation from Newtonian
Mechanics, \textit{Phys. Rev.}, \textbf{vol.~150}, 1966,
pp.1079--1085.
\bibitem{8}J.~Marciak-Kozlowska, Picosecond Thermal Pulses in Thin
Gold Film, \textit{Int. J. Thermophysics}, \textbf{vol.~16}, 1995,
pp.1489-1497.
\bibitem{9}J.~Marciak-Kozlowska, M.~Kozlowski, Velocity of the
Thermal Waves Generated by Femtosecond Laser Pulses in Nanometer
Thick Gold Films, \textit{Lasers in Engineering}, \textbf{vol.~5},
1996, pp.51-58.
\bibitem{10}M.~Dresher et al., \textit{Nature}, \textbf{419},
2002, p.~803.
\end{thebibliography}
\end{document}